\input epsf

\magnification\magstephalf
\overfullrule 0pt
\openup 1.65pt
\def\gsim{\raise.3ex\hbox{$\;>$\kern-.75em\lower1ex\hbox{$\sim$}$\;$}}

\font\rfont=cmr10 at 10 true pt
\def\ref#1{$^{\hbox{\rfont {[#1]}}}$}


\font\tenbfit=cmbxti10
\font\sevenbfit=cmbxti10 at 7pt
\font\fivebfit=cmbxti10 at 5pt
\newfam\bfitfam 
\textfont\bfitfam=\tenbfit  \scriptfont\bfitfam=\sevenbfit
\scriptscriptfont\bfitfam=\fivebfit

\font\tenbfit=cmbxti10
\font\sevenbfit=cmbxti10 at 7pt
\font\fivebfit=cmbxti10 at 5pt
\newfam\bfitfam 
\textfont\bfitfam=\tenbfit  \scriptfont\bfitfam=\sevenbfit
\scriptscriptfont\bfitfam=\fivebfit

\font\tenbit=cmmib10
\newfam\bitfam
\textfont\bitfam=\tenbit%

\font\tenmbf=cmbx10
\font\sevenmbf=cmbx7
\font\fivembf=cmbx5
\newfam\mbffam
\textfont\mbffam=\tenmbf \scriptfont\mbffam=\sevenmbf
\scriptscriptfont\mbffam=\fivembf

\font\tenbsy=cmbsy10
\newfam\bsyfam 
\textfont\bsyfam=\tenbsy%


\def\pmb#1{\setbox0=\hbox{#1}
 \kern.05em\copy0\kern-\wd0 \kern-.025em\raise.0433em\box0 }

\def\slash{/\kern-.5em}


 %


\def\boxit#1{\vbox{\hrule\hbox{\vrule\kern1pt\vbox
{\kern1pt#1\kern1pt}\kern1pt\vrule}\hrule}}

\def\h{\hfill\break}
\parskip=8pt
\parindent=0pt
\hsize=17truecm\hoffset=-5truemm
\vsize=23.5truecm\voffset=-10 truemm
\def\footnoterule{\kern-3pt
\hrule width 17truecm \kern 2.6pt}


\catcode`\@=11 

\def\nolabels{\def\wrlabeL##1{}\def\eqlabeL##1{}\def\reflabeL##1{}}
\def\writelabels{\def\wrlabeL##1{\leavevmode\vadjust{\rlap{\smash%
{\line{{\escapechar=` \hfill\rlap{\sevenrm\hskip.03in\string##1}}}}}}}%
\def\eqlabeL##1{{\escapechar-1\rlap{\sevenrm\hskip.05in\string##1}}}%
\def\reflabeL##1{\noexpand\llap{\noexpand\sevenrm\string\string\string##1}}}
\nolabels
\global\newcount\refno \global\refno=1
\newwrite\rfile
\def\defref{$^{{\hbox{\rfont [\the\refno]}}}$\nref}
\def\nref#1{\xdef#1{\the\refno}\writedef{#1\leftbracket#1}%
\ifnum\refno=1\immediate\openout\rfile=refs.tmp\fi
\global\advance\refno by1\chardef\wfile=\rfile\immediate
\write\rfile{\noexpand\item{#1\ }\reflabeL{#1\hskip.31in}\pctsign}\findarg}
\def\findarg#1#{\begingroup\obeylines\newlinechar=`\^^M\pass@rg}
{\obeylines\gdef\pass@rg#1{\writ@line\relax #1^^M\hbox{}^^M}%
\gdef\writ@line#1^^M{\expandafter\toks0\expandafter{\striprel@x #1}%
\edef\next{\the\toks0}\ifx\next\em@rk\let\next=\endgroup\else\ifx\next\empty%
\else\immediate\write\wfile{\the\toks0}\fi\let\next=\writ@line\fi\next\relax}}
\def\striprel@x#1{} \def\em@rk{\hbox{}} 
\def\lref{\begingroup\obeylines\lr@f}
\def\lr@f#1#2{\gdef#1{\defref#1{#2}}\endgroup\unskip}
\newwrite\lfile
{\escapechar-1\xdef\pctsign{\string\%}\xdef\leftbracket{\string\{}
\xdef\rightbracket{\string\}}}

\def\writestop{\def\writestoppt{\immediate\write\lfile{\string\p
ageno%
\the\pageno\string\startrefs\leftbracket\the\refno\rightbracket%
\string\def\string\secsym\leftbracket\secsym\rightbracket%
\string\secno\the\secno\string\meqno\the\meqno}\immediate\closeout\lfile}}
\def\writestoppt{}\def\writedef#1{}
\catcode`\@=12 
\font\helv=phvr8r scaled 975
\font\bold=phvb8r scaled 975
\font\oblique=phvro8r scaled 975
\def \q {{\bf q}}
\def \P {{\bf P}}
\helv

\rightline{DAMTP-2017-10}
\rightline{MAN/HEP/2017/05}
\centerline{{\bold Perturbative evolution: a different approach at 
small \pmb{$x$}\ }\footnote{*}{partially supported by STFC grant ST/L000385/1}}
\vskip 8pt
\centerline{A Donnachie}
\centerline{School of Physics and Astromony, University of Manchester}
\vskip 5pt
\centerline{P V Landshoff}
\centerline{DAMTP, Cambridge University
\footnote{\dag}{email addresses: Sandy.Donnachie@manchester.ac.uk, \ 
pvl@damtp.cam.ac.uk}}
\bigskip
{\bold Abstract}

We propose an approach to DGLAP evolution at small $x$ that circumvents the usual
problem that a perturbation expansion is not valid there. The data for the charm structure
function are important to motivate the method, and it describes them much more 
successfully than the conventional approach.

{\bold 1 Introduction}

Although fixed-order perturbative QCD is widely applied to data for the proton
structure function $F_2(x,Q^2)$, it is well-known\defref\smallx{
S Catani and F Hautmann, Nuclear Physics B427 (1994) 475 -- arXiv:hep-ph/9405388\h
J R Cudell, A Donnachie and P V Landshoff, Physics Letters B448 (1999) 281 and B526 (2002) 413\h
G Altarelli, R D  Ball and S Forte, Nuclear Physics B599 (2001) 383\h
A A  Almasy, N A  Lo Presti and A  Vogt, JHEP 01 (2016) 028 -- arXiv:1511.08612
}\defref\book{
A Donnachie, H G Dosch, P V Landshoff and O Nachtmann, {\oblique Pomeron Physics
and QCD}, Cambridge University Press (2002)
}
that there are theoretical problems at small $x$. In this paper,
we refine our previous approach\defref\evol{
A Donnachie and P V Landshoff, Physics Letters B533 (2002) 277 -- arXiv:hep-ph/0111427
}
to overcoming this problem, which combines the DGLAP evolution equation 
with Regge theory. 

An important motivation is provided by the HERA data\defref\charmdata
{
H1 and ZEUS collaborations: H Abramowicz et al,  European Physics Journal C73 (2013) 2311 
-- arXiv:1211.1182 and https://www.desy.de/h1zeus/combined\_results/index.php
}
for charm electroproduction, which have $F_2^{c\bar c}(x, Q^2)$ behaving at small $x$
as the same fixed power $x^{-\epsilon_0}$, with $\epsilon_0$ close to 0.4, down to 
very small $Q^2$. This is seen in figure 1. We show only data having $y<0.55$, so
that although strictly the data are for the reduced cross section, this differs
little from $F_2^{c\bar c}(x, Q^2)$. 
The lines in figure 1 are the result of the 
calculations we describe in this paper of $F_2^{c\bar c}(x, Q^2)$ to nonleading
order (NLO) in perturbative QCD. They are to be contrasted with 
calculations\defref\lhapdf
{
https://www.hepforge.org/archive/lhapdf/pdfsets/6.1/ and
A Buckley et al, European Physics Journal C75 (2015) 132 -- arXiv:1412.7420)
}
in the conventional approach to perturbative evolution, two of which are shown in 
figure~2. The conventional approach ignores the need for resummation at small $x$, 
which we do not know how properly to do and which we circumvent.
If anything, the photoproduction data in figure 1b
call for a rise even steeper than $(W^2)^{0.4}$, though the line in the figure does not
include threshold effects, which will pull it down a little at the lower energies.

Data for high energy $pp$ and $\bar pp$ elastic scattering and total cross sections
have long been known to be dominated by so-called soft pomeron exchange, giving a
power behaviour $s^{\epsilon_1}$ to  the amplitude. First fits\defref\total1{
A Donnachie and P V Landshoff, Nuclear Physics B231 (1984) 189; Physics Letters B296 (1992) 227
}
gave a value for $\epsilon_1$ close to 0.08, but more recent data increase
this\defref\cudell{
J R Cudell, K Kang and S K Kim, Physics Letters B395 (1997) 311
}\defref\totaltwo{
A Donnachie and P V Landshoff, Physics Letters B727 (2013) 500 -- arXiv:1309.1292
}
to 0.096.  
The emergence from HERA of data for $F_2(x,Q^2)$
revealed\defref\twopom
{
A Donnachie and P V Landshoff, Physics Letters B437 (1998) 408 
-- arXiv:hep-ph/9806344
}
that at small $x$ the power $x^{-\epsilon_1}$ alone did not fit and
that a further term behaving as $x^{-\epsilon_0}$ needed to be added. The value of
$\epsilon_0$ is about 0.4, but with quite a large uncertainty, perhaps 10\%.
For the complete $F_2(x,Q^2)$ both terms are needed to describe the data, but we have 
seen in figure 1 that for the charm component of $F_2(x,Q^2)$ the ``hard-pomeron''
term $x^{-\epsilon_0}$ is dominant, with extremely little room for the addition
of the ``soft-pomeron'' term $x^{-\epsilon_1}$.

In leading order perturbative QCD charm electroproduction is calculated from the
gluon density $xg(x,Q^2)$, and to a very good approximation this is true also in 
next-to-leading order. Hence $xg(x,Q^2)$ is dominated by hard-pomeron exchange
at small $x$: it behaves as $x^{-0.4}$ even at quite small $Q^2$:
$$
xg(x,Q^2)=G(Q^2)~x^{-\epsilon_0}
\eqno(1)
$$

If we Mellin transform the quark and gluon densities by applying $\int
_0^1dx\,x^{N-1}$,
a power $u(Q^2)x^{-\epsilon}$ in the $u$-quark density, for example, becomes
a pole $u(Q^2)/(N-\epsilon)$.
Write
$$
\q(Q^2)=\left [\matrix{\Sigma(Q^2\cr G(Q^2)\cr}\right ] ~~~~~~~
\Sigma(Q^2)= u(Q^2)+d(Q^2)+s(Q^2)+c(Q^2)
+ \bar u(Q^2)+\bar d(Q^2)+\bar s(Q^2)+\bar c(Q^2)
\eqno(2a)
$$
then
$$
Q^2{d\over{d Q^2}}\q (Q^2)= {{\alpha_s(Q^2)}\over{2\pi}}\P (N=\epsilon,\alpha_s(Q^2))~\q(Q^2)
\eqno(2b)
$$
At nonleading order (NLO) the evolution matrix is 
$$
\P (N,\alpha_s)=\P^0 (N)+{\alpha_s\over{2\pi}} \, \P^1 (N)
\eqno(2c)
$$
The problem with applying this to perturbative evolution at small $x$ lies in
that, while the $qq$ and $qg$ elements of $\P^0 (N)$ are finite at $N=0$, its other two
elements, and those of $\P^1 (N)$, have poles there\defref\esw{
R K Ellis, W J Stirling and B R Webber, {\oblique QCD and Collider Physics}, Cambridge University Press (1996)
}.
This means that at $N=0.096$ the NLO evolution matrix is very different from its 
leading order (LO) version.

Of the four elements of the evolution matrix $\P(N)$, it is\ref{\esw}
$P_{qg}$ that differs most between LO and NLO. 
As is seen in Figure 3a, for $N$ between 0.096 and 0.4 good fits to $P^0_{qg}(N)$ and $P^1_{qg}(N)$ are
$$
P^0_{qg}(N)=2.59-2.21N+1.085N^2
$$$$
P^1_{qg}(N)=24.6/N-28.2
\eqno(3a)
$$
As a model for the exact $P_{qg}(N)$ consider
$$
P_{qg}=P^0_{qg}(N)+C\sqrt{C^2N^2+N\alpha_s P^1_{qg}(N)/\pi}-C^2N
$$
$$
~~~~~~~~~~~~~~~~~~=P^0_{qg}(N)+\alpha_s P^1_{qg}(N)/(2\pi)-\alpha_s^2 [P^1_{qg}(N)]^2/(8\pi^2NC^2)+\dots
\eqno(3b)
$$
where $C$ is a constant.
Figure 3b shows  the exact $P_{qg}$ in this model, together with the NLO and
NNLO approximations to it.  The conclusion from this model is that, provided the 
NLO and the NNLO approximations are
not very different, then NLO is already a good approximation. Further, this
is much more likely to be true for $N=0.4$ than for $N=0.096$.

The problem with the perturbative expansion at small $x$
is implicit in all conventional applications of the DGLAP equation:
for values of $Q^2$ for which the PDFs are fairly flat in $x$ at small $x$,
that is the effective power $(1/x)^{\epsilon}$ is small, it
is not valid to use the DGLAP equation  with $\P (N,\alpha_s(Q^2))$
expanded in powers of $\alpha_s(Q^2)$.

But we have inferred that the contribution from soft pomeron
exchange to the gluon density is negligibly small.  
So we apply NLO perturbative QCD evolution (2b) with $N=\epsilon_0$
to the coefficient of the
hard pomeron contribution to $F_2(x,Q^2)$. We have seen in figure 1 that the resulting 
gluon density leads to a highly successful calculation of $F_2^{c\bar c}(x,Q^2)$ to
NLO. Figure 4 shows data \defref\zeusfl
{
ZEUS Collaboration: S Chekanov et al, Physics Letters B682 (2009) 8 --  arXiv:0904.1092
}
\defref\hfl
{
H1 collaboration: F D Aaron et al, Physics Letters B393 (1997) 452 -- arXiv:1012.4355
}
that indicate that the same is true of a calculation to leading order (LO) 
of the longitudinal structure function $F_L(x,Q^2)$. Unlike for $F_2(x,Q^2)$,
the NLO correction is known to be small\defref\moch
{
S Moch, J A M Vermaseren and A Vogt, Physics Letters B606 (2005) 123
}.

\bigskip
{\bold 2 Fit to data for \pmb{$F_2$}}

The H1 and ZEUS experiments at HERA have agreed joint  data
for the proton structure function $F_2(x,Q^2)$. Their 2010 publication\defref\heraone{
H1 and ZEUS collaborations: F D Aaron et al,  JHEP 01 (2010) 109
}
both included results for the so-called reduced cross section
$$
\sigma^{\hbox{{\sevenrm red}}}(x,y,Q^2)=F_2(x,Q^2)-{y^2\over 1+(1-y)^2}F_L(x,Q^2)
\eqno(4)
$$
and an extraction of $F_2(x,Q^2)$ from it, though only for events with $y$ less than
about 0.5. Their 2015 publication\defref\heratwo{
H1 and ZEUS collaborations: H Abramovicz et al, arXiv:1506.06042
}
gave results only for $\sigma^{\hbox{{\sevenrm red}}}(x,y,Q^2)$ and did not attempt to extract $F_2(x,Q^2)$.

We fit the small-$x$ 2010 HERA data\ref{\heraone} for $F_2(x,Q^2)$ to a sum of three powers of $x$:
$$
F_2(x,Q^2)=A_0(Q^2)x^{-\epsilon_0}+A_1(Q^2)x^{-\epsilon_1}+A_2(Q^2)x^{-\epsilon_2}
\eqno(5)
$$
Here, the first two terms are hard and soft pomeron exchange. The last term is 
reggeon exchange, that is the families $f_2$ and $a_2$;  its contribution for the
small values of $x$ at which we work is small.

The only theoretical constraint we know on the three coefficient functions $A_i(Q^2)$ is
that, as $Q^2\to 0$, $F_2(x,Q^2)$ vanishes linearly in $Q^2$ at fixed $2\nu=Q^2/x$. 
We respect this constraint, even though our fit is for values of $Q^2$ some way
from zero. This is sufficient for our purpose, because perturbative evolution is valid
only for ''large'' values of $Q^2$. The theory does not tell how large but we shall find
that $Q^2$ must be at least about 5~GeV$^2$.

We have tried various forms of the $A_i(Q^2)$. Several
give much the same value of $\chi^2$ in the fit to the data, but rather different values
of $\epsilon_0$, which is why there is still a rather large uncertainty in its value.
In this paper we give the results for the choices
$$
A_i(Q^2)=X_i\,(Q^2/(Q_i^2+Q^2))^{1+\epsilon_i}\,(1+2Q^2/Q_i^2)^{0.15}
~~~~~~i=0,1,2
\eqno(6)
$$

We use the values
$$
\epsilon_0=0.4~~~~~~~~~\epsilon_1=0.096~~~~~~~~~\epsilon_2=-0.343
\eqno(7)
$$
where the last two are
extracted\ref{\totaltwo} from $pp$ and $\bar pp$ elastic scattering and total cross
section data.

We confine
attention to the small-$x$ data, which we take to mean $x<10^{-3}$. 
But we introduce into each term a power of $(1-x)$ given by the dimensional
counting rule\defref\roberts{
R G Roberts, {\oblique The Structure of the Proton}, Cambridge University
Press (1990)
}, that is $(1-x)^7$ for each of the two pomeron terms, and $(1-x)^3$ for the reggeon
term. Although this is not strictly correct, it is better than not doing it, at least if $Q^2$ is
not too large. Its effect is small for small $x$, though not completely negligible, and
it makes the fit to the $x<0.001$ data valid some way beyond that range and
therefore to larger $Q^2$: see figure 5.
The fits in this figure have
\vfill\eject
$$
X_0=0.03527~~~  X_1=0.03786~~~  X2=0.06445
$$
\vskip -7 truemm
$$
  Q_0^2=7.03262~~~ Q_1^2=1.68993\times 10^{-5}~~~  Q_2^2=0.00192
\eqno(8)
$$
The $\chi^2$ per data point is 0.95 for all the data 
with $Q^2$ between 3.5 and 45~GeV$^2$ and $x<0.001$. 
\bigskip
{\bold 3 Perturbative QCD evolution}

We work to next-to-leading order in $\overline{MS}$ perturbative QCD, with\defref\pdg{
Particle Data Group, http://pdg.lbl.gov
}
$$
\alpha_S(M_Z^2)=0.1183
\eqno(9a)
$$
We follow the standard procedure\ref{\esw}
of changing from 4 to 5 flavours at $Q^2=m_{b0}^2$
and making $\alpha_S(Q^2)$ continuous there, resulting in
$$
\Lambda_5=230\hbox{ MeV}~~~~\Lambda_4=328\hbox{ MeV}
\eqno(9b)
$$

For the range of $Q^2$ values that will concern us the contribution to
$F_2(x,Q^2)$ from $b$ quarks is\defref\bdata
{
ZEUS collaboration: H Abramowicz et al, JHEP 09 (2014) 127
-- arXiv:1405.6915\h
H1 collaboration: A Aktas et al, European Physics Journal C40 (2005) 349
-- arXiv:hep-ex/0411046; F D Aaron et al, European Physics Journal C65 2010) 89
-- arXiv:0907.2643
}
negligibly small.
Pomerons couple equally to $u$ and $d$ quarks.
In the case of the soft pomeron, we know\ref{\total} that the coupling to $s$ quarks
is rather weaker than to $u$ and
$d$ and, as may be deduced from the data in figure~2,
very much so to $c$. However, the data for $F_2^{c\bar c}(x,Q^2)$ indicate that,
for reasons that are not understood and to a very good approximation, the
hard pomeron couples equally to all four quarks. To verify this, note that
if this is the case,
$F_2^{c\bar c}(x,Q^2)\sim 0.4~A_0(Q^2)x^{-\epsilon_0}$ at small $x$,
and the lines in figure~1, while calculated by NLO perturbative QCD,  
correspond to this to a very good approximation.
As we have said above, the gluon density must also behave as
$x^{-\epsilon_0}$ at small $x$.

In order to apply DGLAP evolution, according to (1) we need$^*$\footnote{}
{$^*$ We
are grateful to the authors of reference {\esw} for making available to us the
numerical values of the elements of $\P (N,\alpha_s(Q^2))$}
the elements of the evolution matrix (2c) at $N=0.4$. We find that, for values of 
$N$ close to this, good fits are
$$
\P ^0(N)=\left [\matrix{-0.0352299-2.41822N+0.743989N^2&
2.58893-2.2144N+1.08471N^2\cr
18.4131-49.7311N+42.476N^2&
39.0374-115.451N+96.1001N^2\cr}\right]
$$$$
\P ^1(N)=\left [\matrix{-1111.55-3187.40N+-2685.19N^2&  
-943.22+2952.70N-2460.29N^2\cr
-636.76+2070.57N-1707.29N^2&  
-114.61+383.00N-359.98N^2\cr}\right]
\eqno(10)
$$ 

We choose some value $\bar Q^2$ of $Q^2$ and calculate $\Sigma(\bar Q^2)$ and
its derivative $\Sigma'(\bar Q^2)$ using the form of $A_0(Q^2)$ we obtained from the fit 
(6) to $F_2(Q^2)$. The DGLAP equation
(2b) with $N=0.4$ then determines $G(\bar Q^2)$. This provides the input to the evolution
in both directions from $Q^2=\bar Q^2$. 
Figure 6 shows the percentage difference for
the choice $\bar Q^2$=9~GeV$^2$. The plot in this figure is quite sensitive to the 
choice of $\bar Q^2$. It is known that perturbative evolution is valid
only at sufficiently large $Q^2$; we conclude that this means greater than about
5~GeV$^2$. Figure 7 shows how $G(Q^2)$ and $\Sigma (Q^2)$
evolve with $Q^2$.  The value $\bar Q^2$=9~GeV~$^2$
gives agreement with the gluon density at very large $Q^2$ obtained from the
conventional approach to perturbative evolution\defref\lhapdf
{
https://www.hepforge.org/archive/lhapdf/pdfsets/6.1/
}
: see figure 8.
At these values of $Q^2$ the effective power $(1/x)^{\epsilon}$ for the behaviour
of $xg(x,Q^2)$ has become so large that the conventional approach is now valid.
\vfill\eject
{\bold 4 Applications}

We now apply the gluon density (1) shown in figure 7 from the NLO calculation
of $G(Q^2)$ using the perturbative evolution formula (1b).
A good fit in the range $5<Q^2<200$~GeV$^2$ is
$$
G(Q^2)=X_g\big (Q^2/(Q_g^2+Q^2)\big )^{\eta_1}\,(1+Q^2/Q_g^2)^{\eta_2}
\eqno(11a)
$$
with
$$
X_g = 0.4338~~~~ \eta_1 = 0.437 ~~~~\eta_2 = 0.252~~~~  Q_g^2 = 9.128
\eqno(11b)
$$

Figure 1a shows the NLO calculation of $F_2^{c\bar c}(x,Q^2)$ using an
adaptation of a program\footnote{*}
{We are grateful to Johannes Bluemlein for making this program available to us}
originally written by
Laenen, Riemersma, Smith and van Neerven\defref\lssvn
{
E Laenen, S Riemersma, J Smith and  W.L. van Neerven, Nuclear Physics B392 (1993) 162
}.
We use the NLO running mass $m_c(Q^2)$ with\ref{\pdg}
$$
m_c(m_{c0})=m_{c0}~~~~~~~~~~~~~~~~~~~m_{c0}=1.27\hbox{ GeV}
\eqno(12a)
$$
A good fit, for $5<Q^2<10^4$~GeV$^2$, is provided by
$$
\big (m_c(Q^2)\big )^2=m_{c0}^2/\big (1+0.3438\log(Q^2/m_{c0}^2)\big )
\eqno(12b)
$$
In both $\alpha_s(Q^2)$ and $xg(x,Q^2)$ we replace $Q^2$ with
$$
\mu^2= Q^2+4\big (m_c(Q^2)\big )^2
\eqno(13)
$$
Our calculation, whose results are shown in figure 1, includes only the 
gluon-induced charm production.
The cross section for charm electroproduction is the sum of two terms, 
$e\,g\to\hbox{charm}$ and $e\,q\to\hbox{charm}$, each convoluted with the corresponding 
parton density. Figure 1 shows that the first alone, calculated up to order $\alpha_s^2$,
fits the data well. For the second, the order $\alpha_s^2$ term is the first term 
in the perturbative expansion. It turns out to be small at small $Q^2$, but it is  
negative because of the inclusion of a counterterm to cancel a collinear 
divergence\footnote{\dag}{We are grateful to Eric Laenen for explaining this to us}, 
and therefore it is a very poor approximation to the exact expression. So it is 
appropriate to neglect it.

We perform an exactly similar calculation of $F_2^{b\bar b}(x,Q^2)$,
with\ref{\pdg}
$$
m_b(m_{b0})=m_{b0}~~~~~~~~~~~~~~~~~~~m_{b0}=4.18\hbox{GeV}
\eqno(14a)
$$
and the fit
$$
(m_b(Q^2))^2=m_{b0}^2/\big (1+0.179\log(Q^2/m_{b0}^2\big )
\eqno(14b)
$$
The resulting calculations are shown in figure 9, together with the data\ref{\bdata}.

We have said that, to calculate the $F_L(x,Q^2)$, it is sufficient to use the
LO formula\defref\budnev
{
V M Budnev et al, Physics Reports C15 (1974) 181
}
$$
F_L(x,Q^2)=K(x,Q^2)+{4\alpha_s(Q^2)\over 3\pi}\int _x^1{dy\over y}
\Big ({x\over y}\Big)^2F_2(y,Q^2)
\eqno(15a)
$$
where the contribution of the charm quark to $K(x,Q^2)$ is
$$
K^c(x,Q^2) = 2e_c^2{\alpha_s(\mu^2)\over \pi}\int _{x\mu^2/Q^2}^1dy\,
\Big ({x\over y}\Big)^2 \Big[\Big(1-{x\over y}\Big)v
-{2m_c^2(Q^2)x\over Q^2y}L\Big)\Big]\,g(y,\mu^2)
\eqno(15b)
$$
with $\mu^2$ defined in (13) and
$$
v^2(Q^2)=1-{{4m_c^2(Q^2)}\over{Q^2(y/x-1)}}
~~~~~~~L=\log\Big ({{1+v}\over{1-v}}\Big )
\eqno(15c)
$$
The light quarks contribute similarly, with zero mass. The calculations are
compared with data\defref\fldata
{
H1 collaboration: F D Aaron et al, C71 (2011) 1579 -- arXiv:1012.4355\h
ZEUS collaboration: S Chekanov et al, Physics Letters B682 (2009) 8 -- arXiv:0904.1092
}
for $F_L(x,Q^2)$ in figure 4. The circles in figure 10 show the calculated points 
for the reduced cross section (4), together with the 2015 joint HERA data\ref{\heratwo}. 
The lines in this figure are the fits to $F_2(x,Q^2)$ from the 2010 HERA 
data\ref{\heraone}. 
The $\chi^2$ per data point is 0.96.  
Although, according to figure 6, the DGLAP evolution is not valid 
below $Q^2=5$~GeV$^2$, the evolved gluon density gives good results for $F_L(x,Q^2)$
down to 3.5~GeV$^2$. 
\bigskip
{\bold Acknowledgments}

We thank Marco Guzzi for helpful discussions on DGLAP and the CT14 analysis.

\medskip\immediate\closeout\rfile\writestoppt
\baselineskip=12pt{{\bold References}}\bigskip{\frenchspacing%
\parindent=20pt\escapechar=` \input refs.tmp\bigskip}\nonfrenchspacing
\pageinsert
\epsfxsize=0.45\hsize\epsfbox[70 50 375 280]{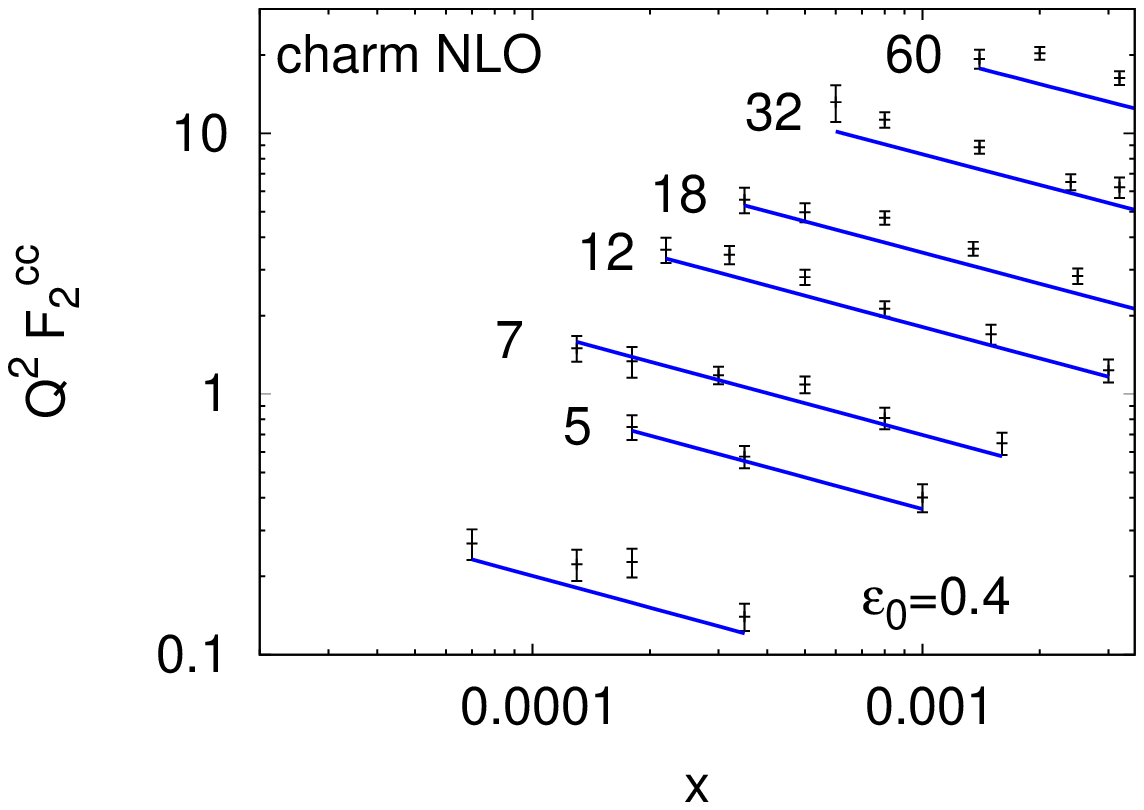}\hfill
\epsfxsize=0.45\hsize\epsfbox[70 50 375 280]{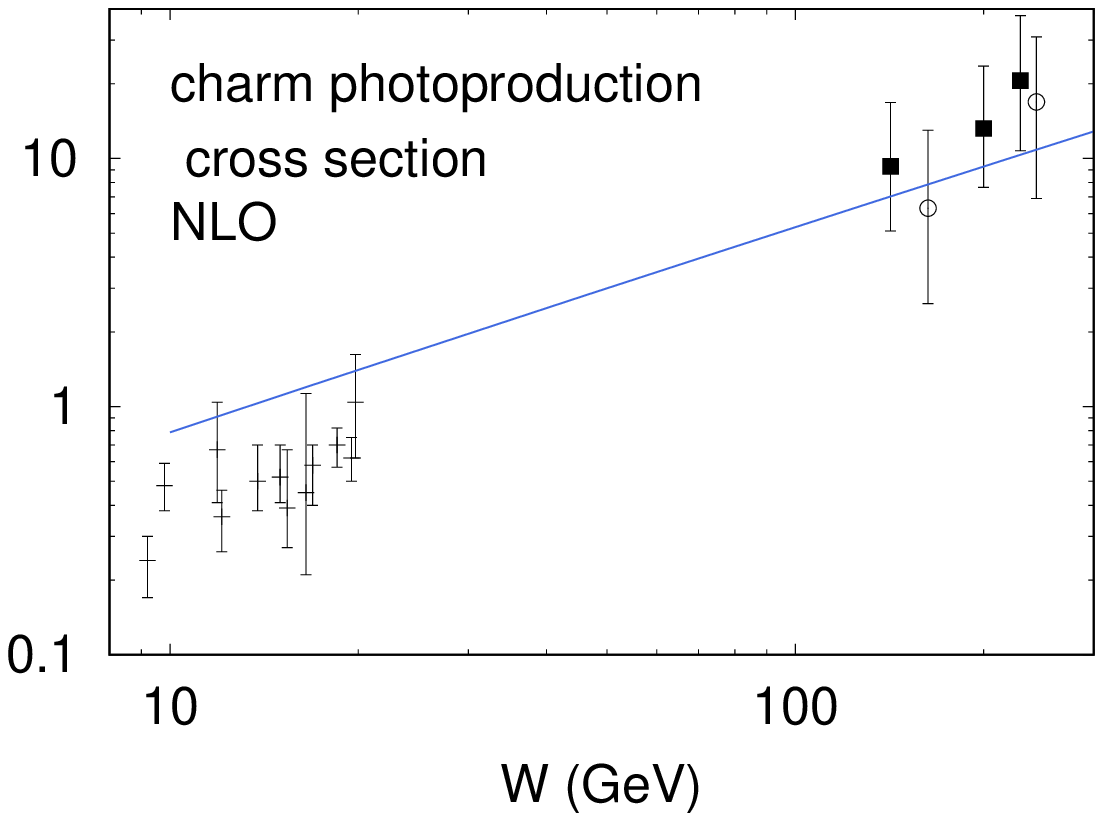}

\centerline{$\phantom{XXX}$(a)$\phantom{XXXXXXXXXXXXXXXXXXXXXXX}$(b)}

Figure 1: (a) Data\ref{\charmdata} for for charm
electroproduction, with lines proportional to $x^{-0.4}$ and (b) for
charm photoproduction compared with $(W^2)^{0.4}$.
\bigskip
\epsfxsize=0.45\hsize\epsfbox[70 60 375 280]{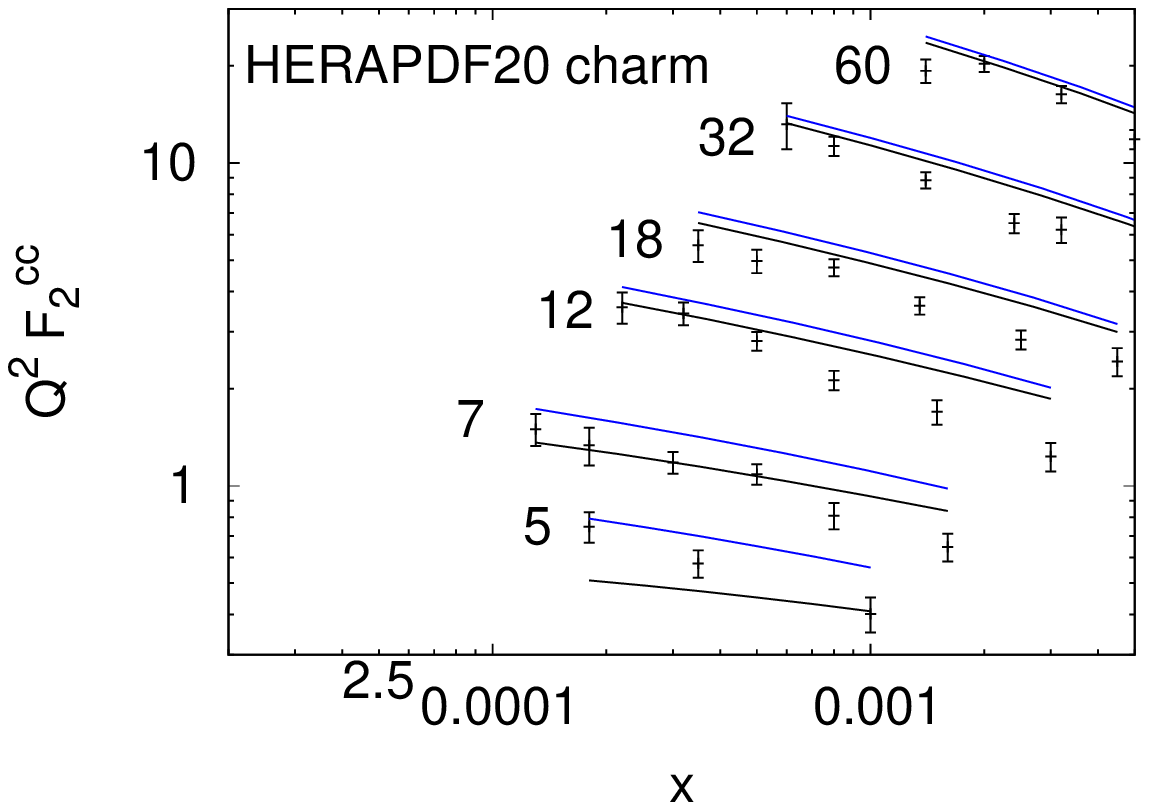}\hfill
\epsfxsize=0.45\hsize\epsfbox[70 60 375 280]{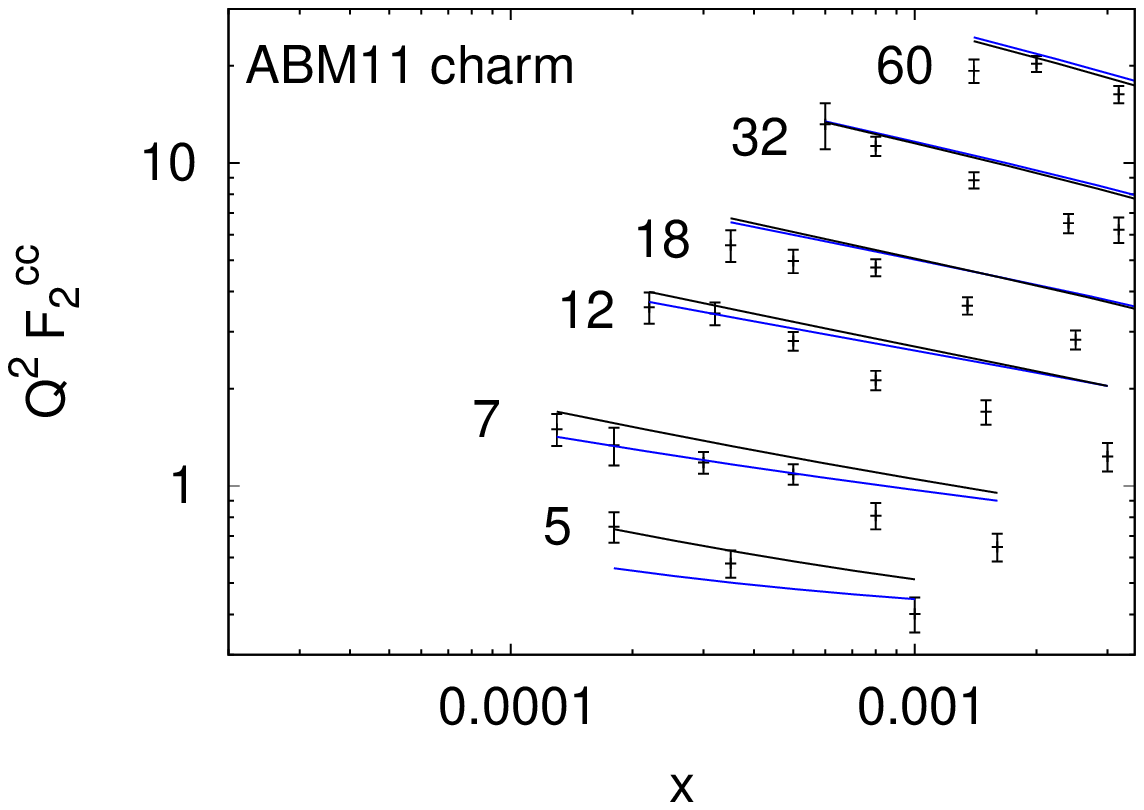}

\centerline{$\phantom{XXX}$(a)$\phantom{XXXXXXXXXXXXXXXXXXXXXXX}$(b)}

Figure 2: Calculations\ref{\lhapdf} of $F_2^{c\bar c}(x,Q^2)$ at NLO and NNLO. 
In (a) the NNLO calculations are below the NLO, while in (b) they are above. 
\bigskip
\epsfxsize=0.4\hsize\epsfbox[70 60 375 280]{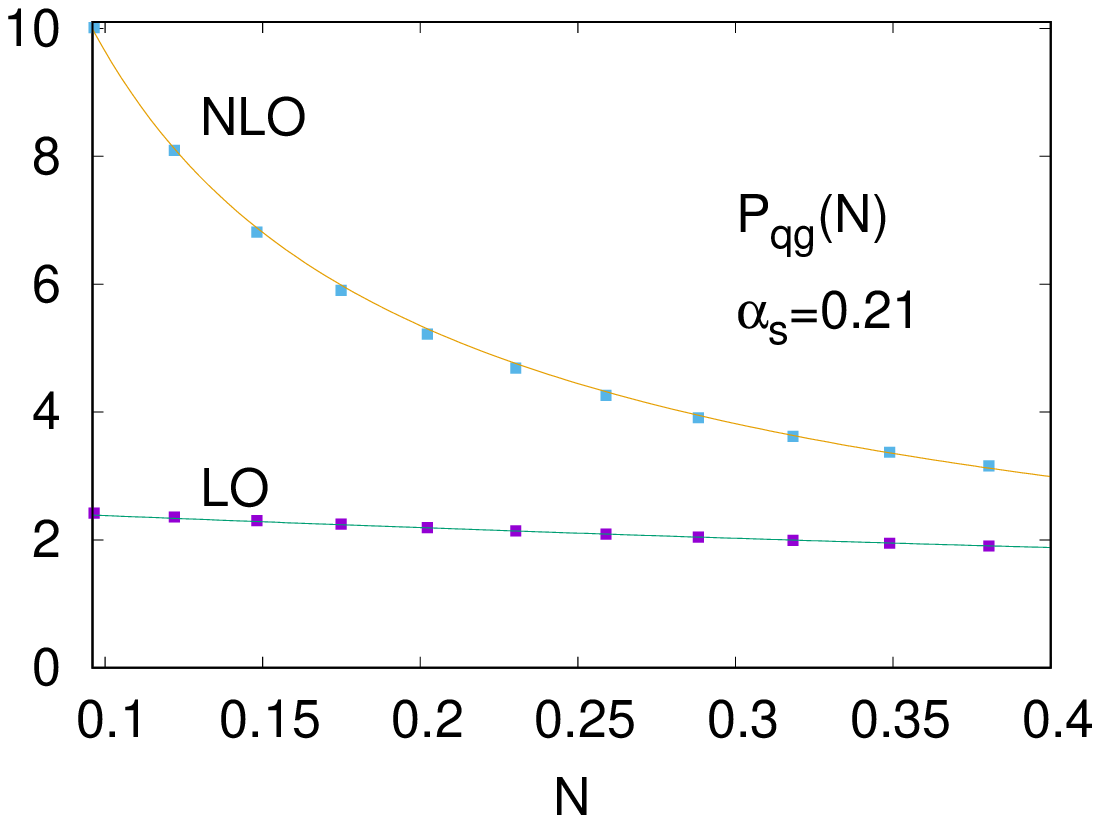}\hfill
\epsfxsize=0.4\hsize\epsfbox[70 60 375 280]{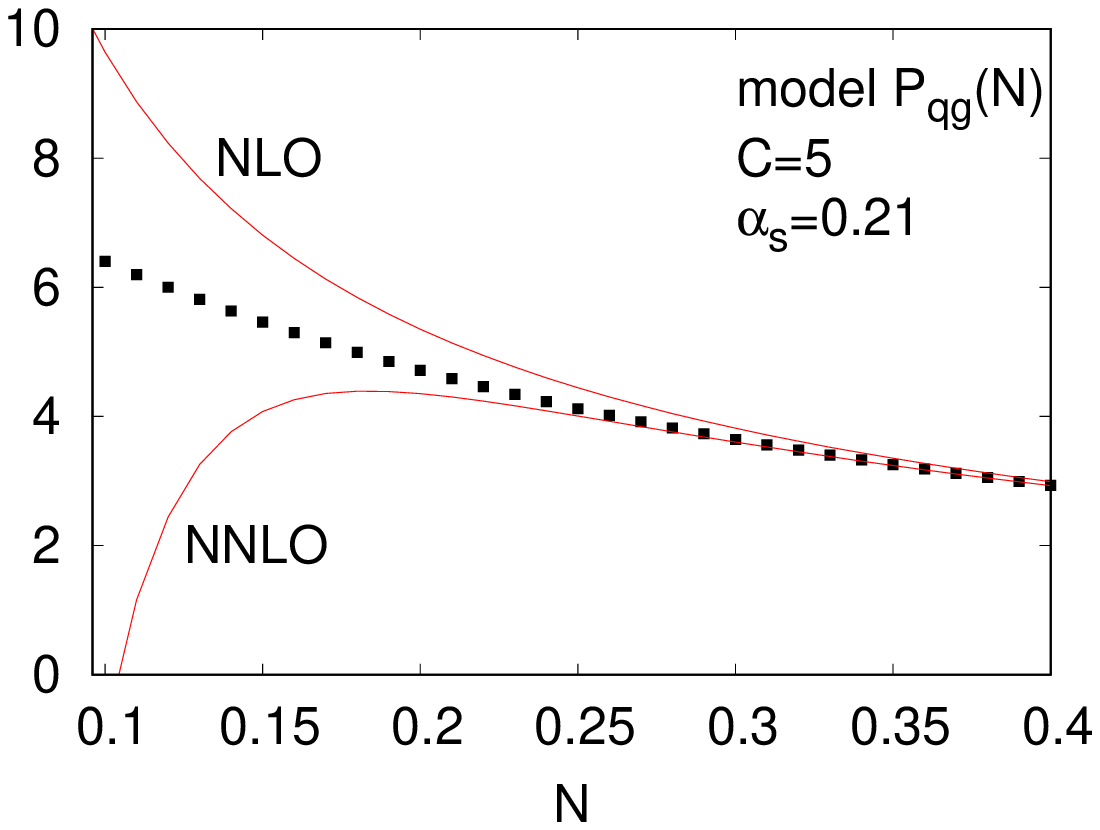}

\centerline{$\phantom{XXX}$(a)$\phantom{XXXXXXXXXXXXXXXXXXXXXXX}$(b)}

Figure 3: (a) The evolution matrix element\ref{\esw}
$P_{qg}$ for $\alpha_s=0.21$ at LO and NLO
together with the fits (3a); (b) the exact $P_{qg}$ in the model 
(3b) (points) with the NLO and NNLO approximations
\endinsert
\pageinsert
\epsfxsize=0.45\hsize\epsfbox[50 50 375 290]{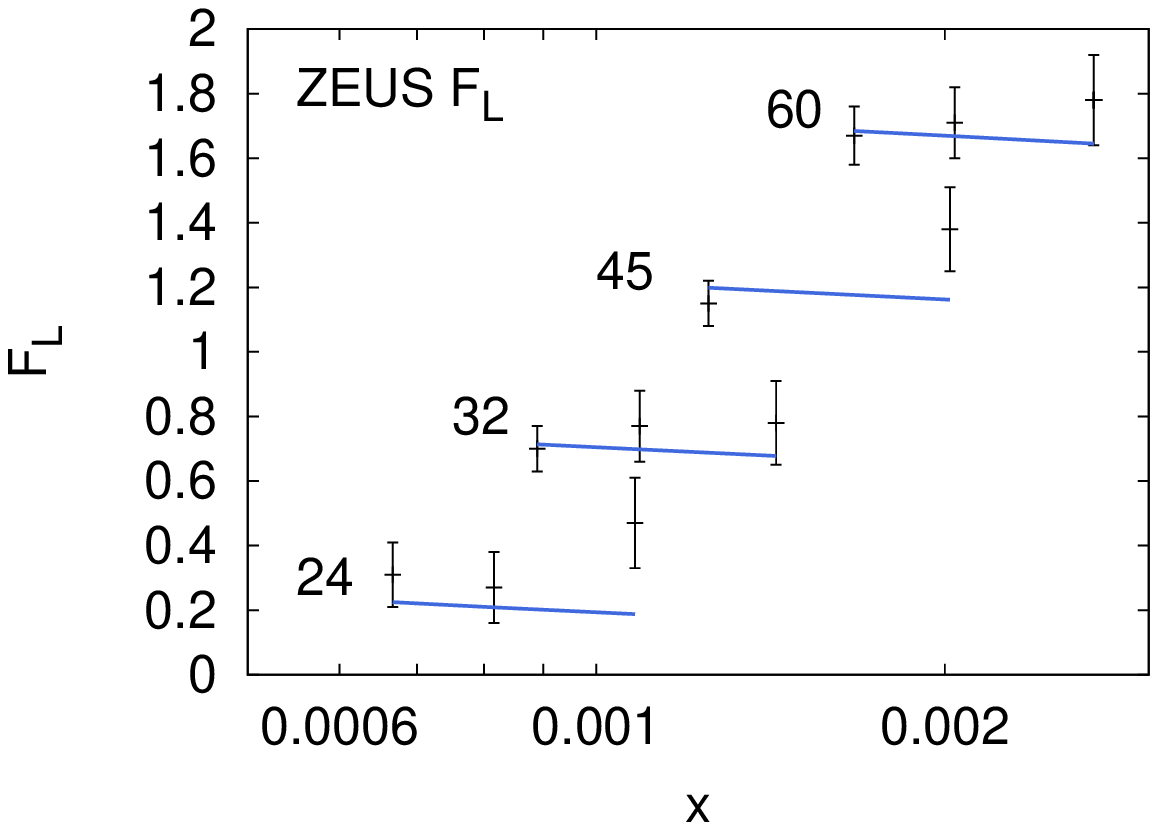}\hfill
\epsfxsize=0.45\hsize\epsfbox[50 50 375 290]{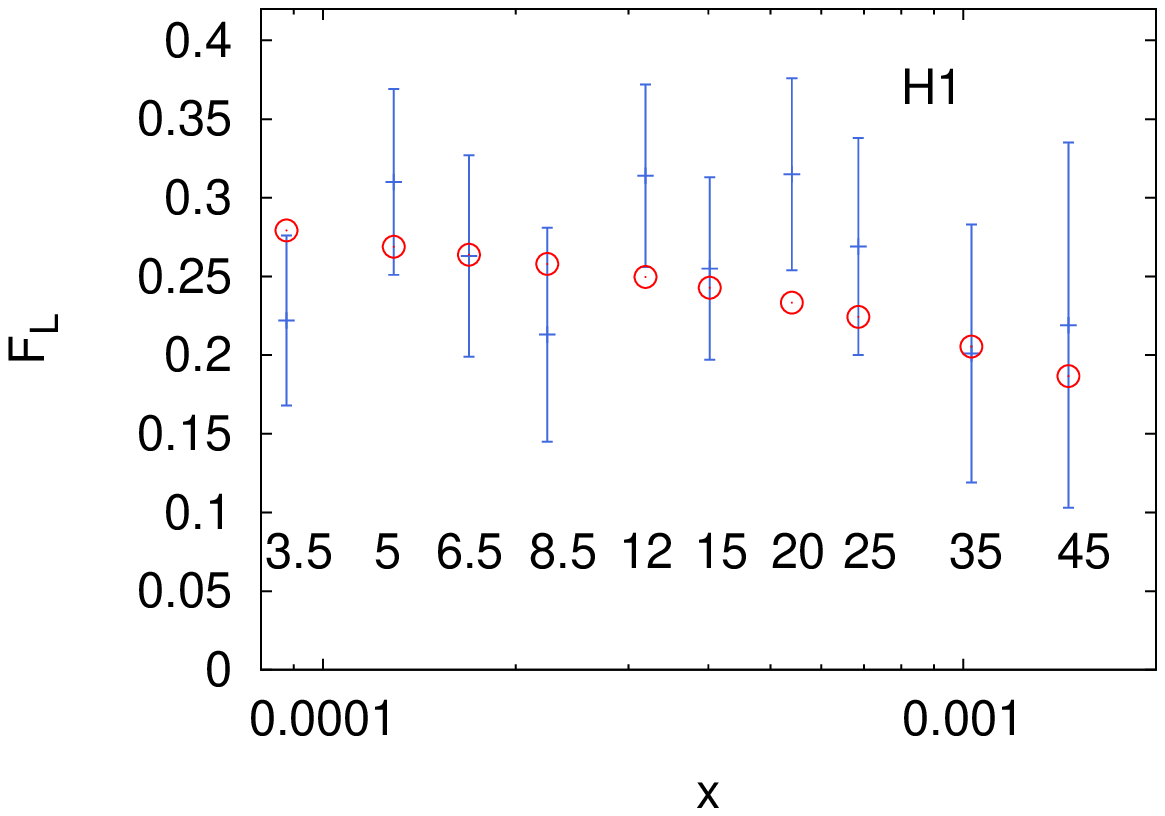}

Figure 4: LO calculation of $F_L(x,Q^2)$ at the indicated values of $Q^2$ compared with 
data from ZEUS\ref{\zeusfl} and H1\ref{\hfl}.
\bigskip
\centerline
{\epsfxsize=0.55\hsize\epsfbox[50 50 375 535]{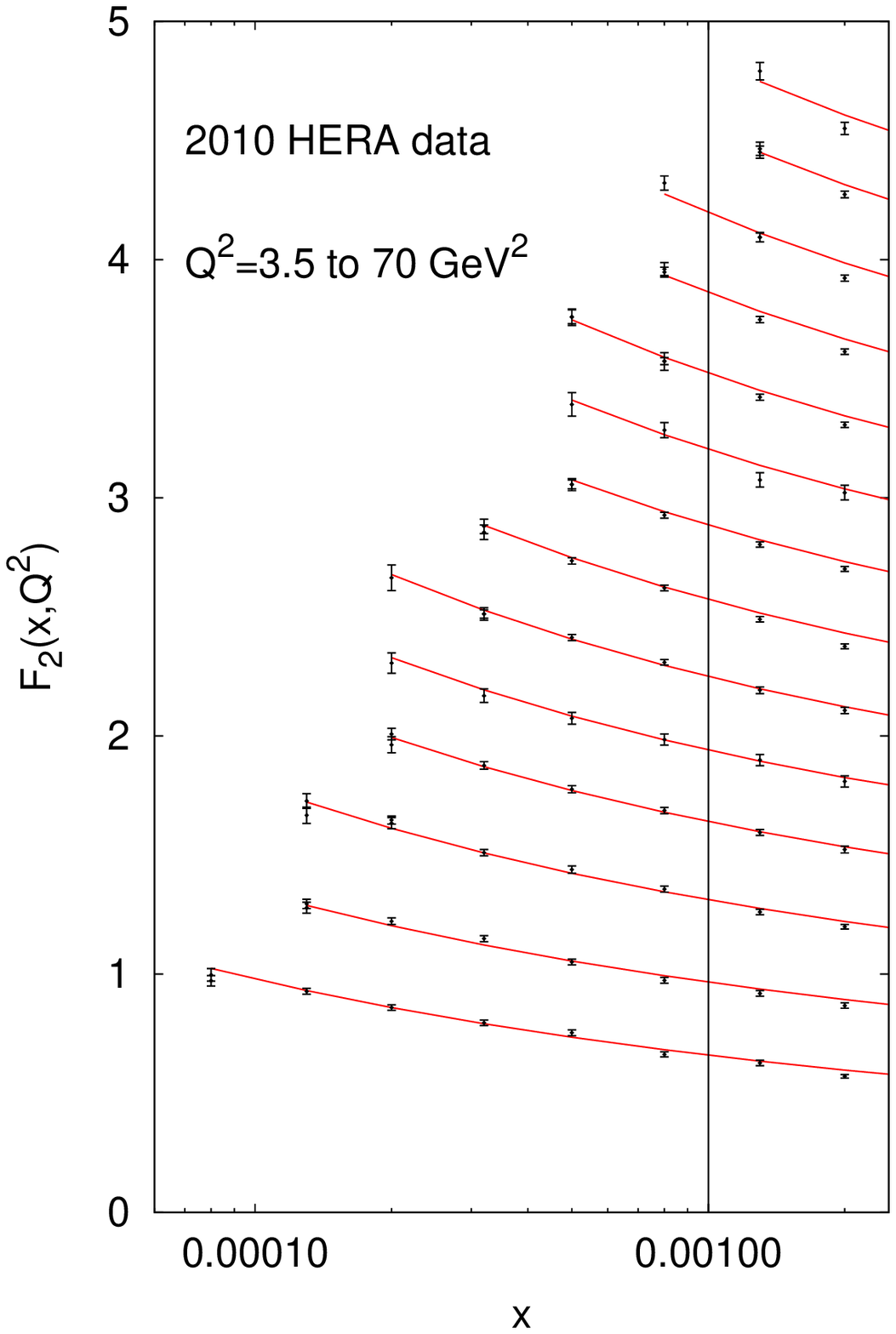}}

Figure 5: fit to 2010 HERA data\ref{\heraone} -- only data with $x<0.001$ are included in the fit
\endinsert
\pageinsert
\centerline{\epsfxsize=0.5\hsize\epsfbox[50 50 375 220]{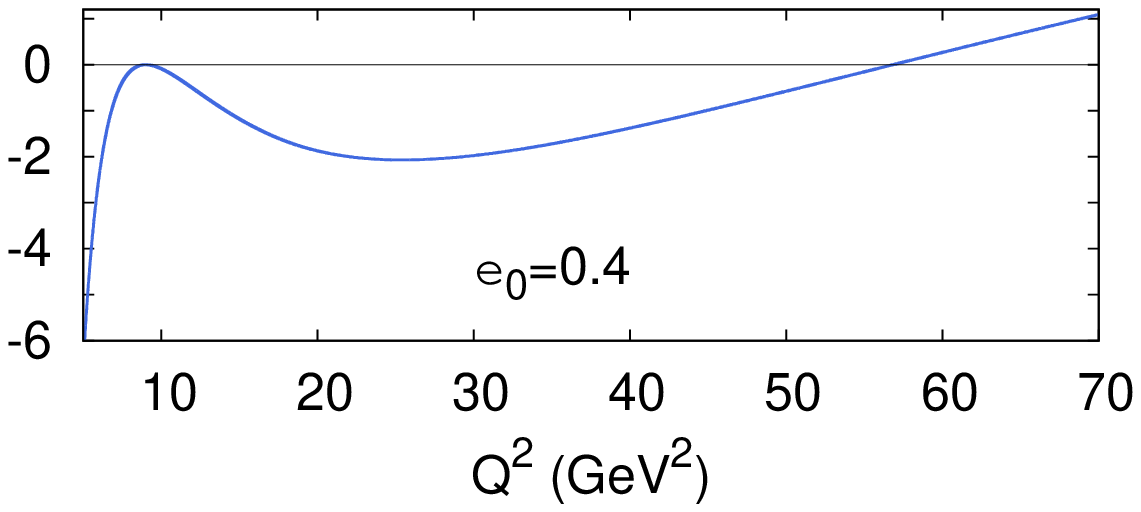}}
\vskip -4 truemm
Figure 6: Percentage difference between the evolved $\Sigma(Q^2)$ and that obtained
from the fit to $A_0(Q^2)$ from the small-$x$ data for $F_2(x,Q^2)$
\bigskip
\centerline{\epsfxsize=0.45\hsize\epsfbox[50 50 375 290]{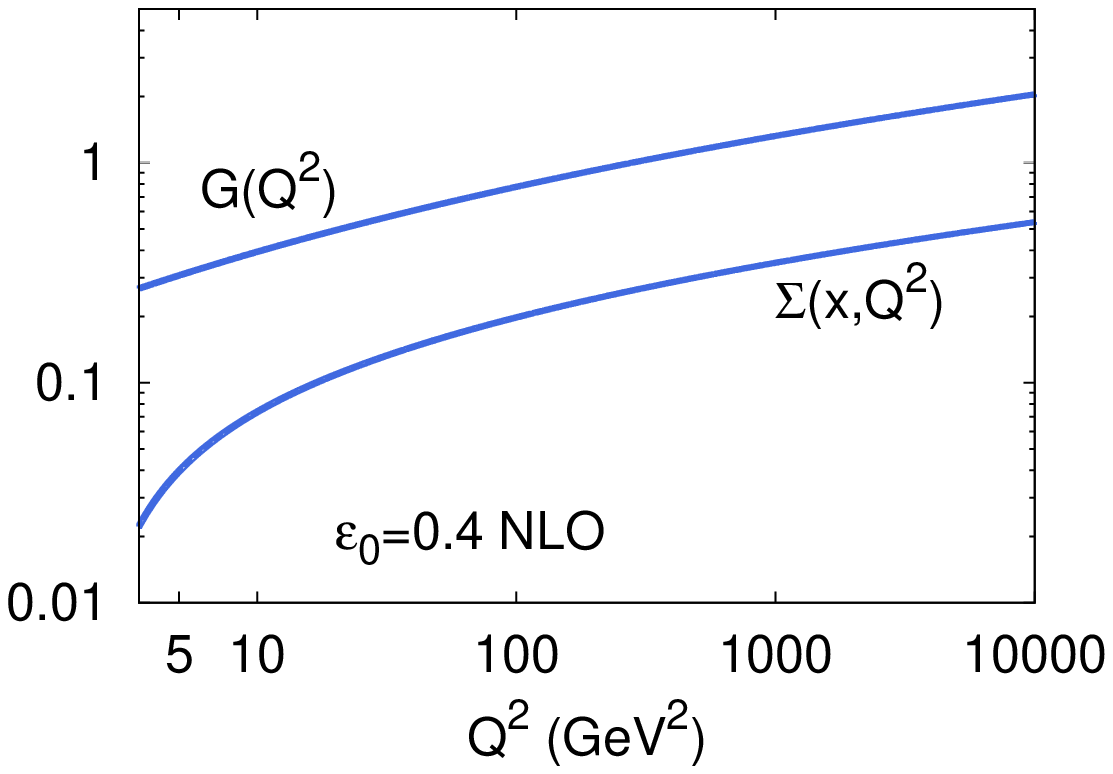}}
\vskip -4 truemm
Figure 7: Evolution of $G(Q^2)$ and $\Sigma (Q^2)$
\bigskip
\centerline{\epsfxsize=0.45\hsize\epsfbox[50 50 375 290]{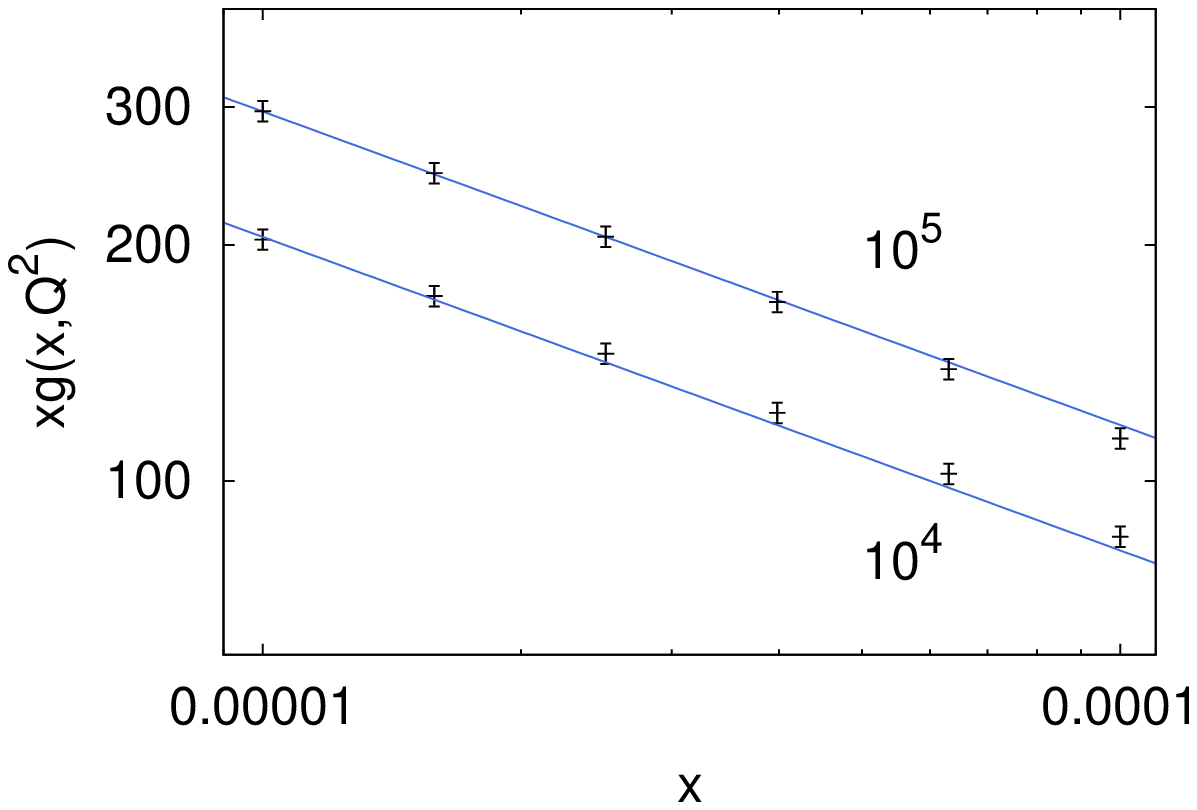}}
\vskip -4 truemm
Figure 8: Evolved gluon density at $Q^2=10^4$ and $10^5$ GeV$^2$.
The points are the results\ref{\lhapdf}
of conventional evolution according to HERAPDF20.
\bigskip
\bigskip
\epsfxsize=0.45\hsize\epsfbox[30 30 375 280]{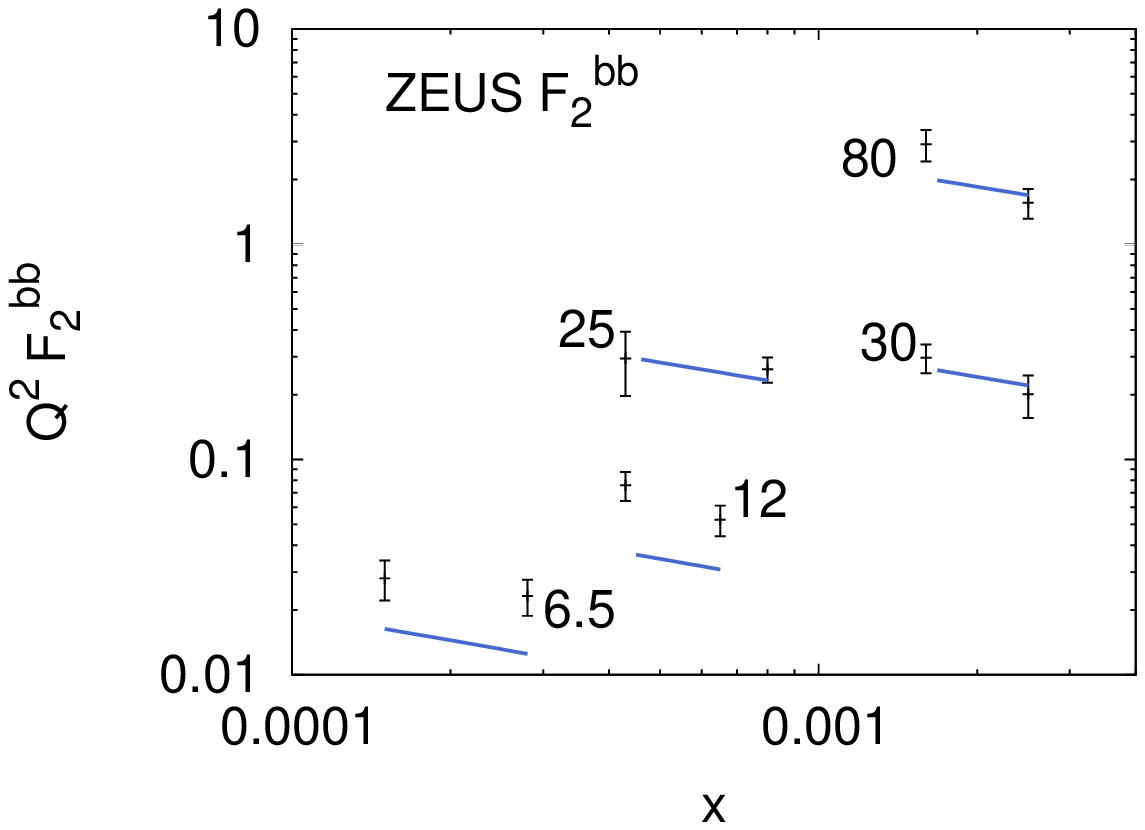}
\hfill
\epsfxsize=0.45\hsize\epsfbox[30 30 375 280]{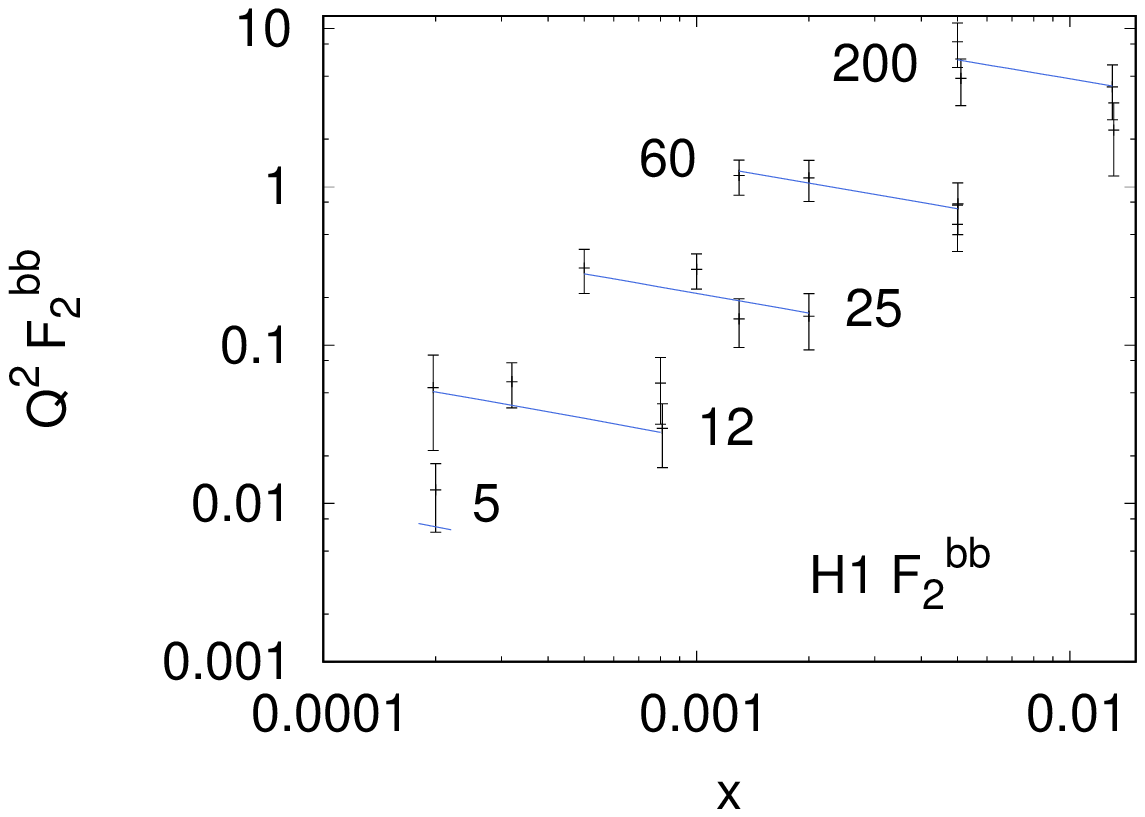}
\vskip -4 truemm
Figure 9: NLO calculations of $F_2^{b\bar b}(x,Q^2)$ with data\ref{\bdata}
 from ZEUS and H1 at various values of $Q^2$.
\endinsert
\pageinsert
\centerline{\epsfysize=0.8\vsize\epsfbox[50 50 375 535]{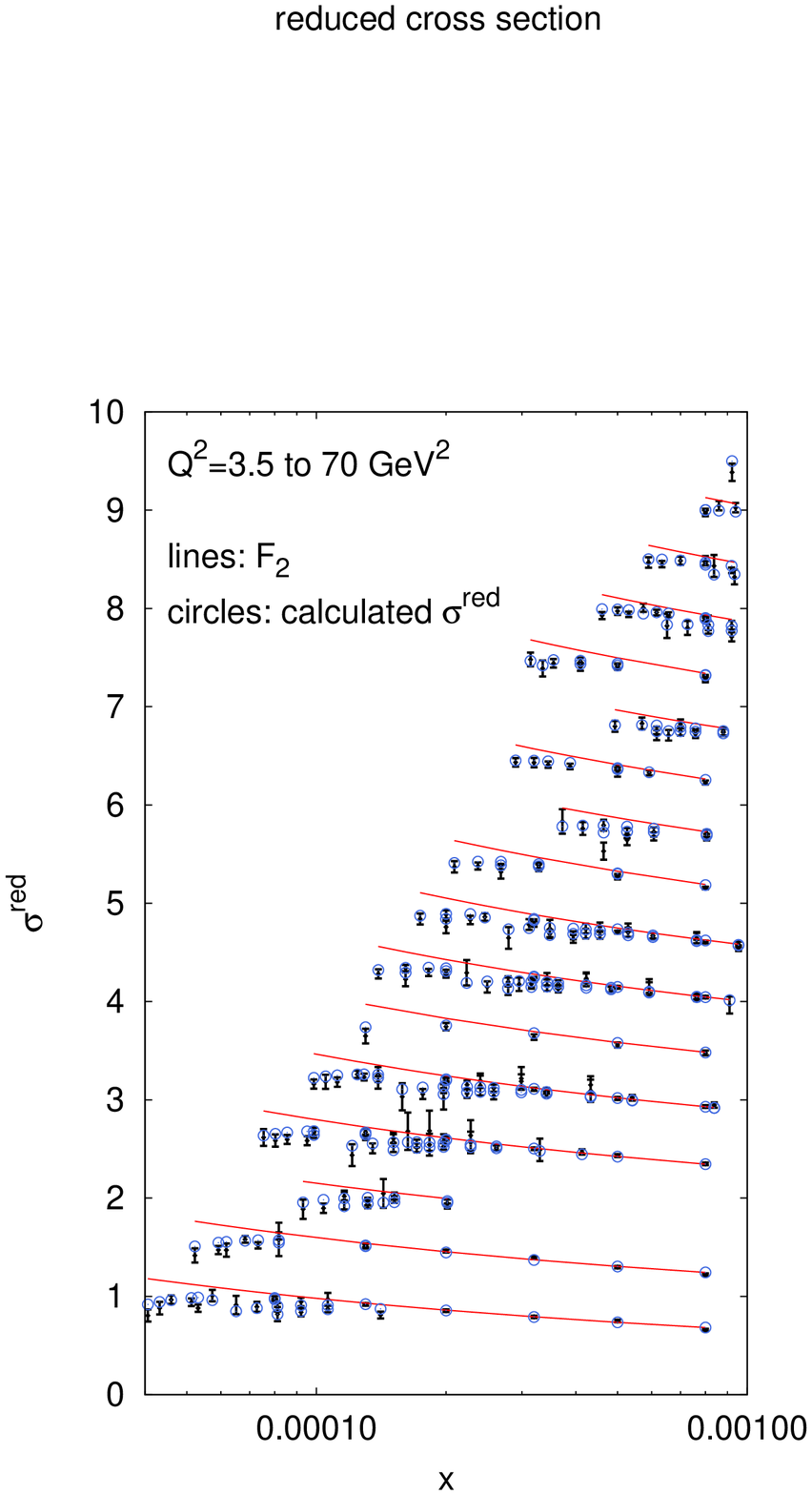}}
\vskip 1pt
Figure 10: Data\ref{\heratwo} for the reduced cross section. The circles denote the 
calculations and the lines the fit to $F_2(x,Q^2)$ from the 2010 data\ref{\heraone}
-- 0.25 is added to the data points for each successive value of $Q^2$.

\endinsert
\bye